**Lead Halide Perovskite based Dynamic Metasurfaces**

*Chen Zhang, Yuhan Wang, Yisheng Gao, Yubin Fan, Can Huang, Nan Zhang, Wenhong Yang, Qinghai Song\*, Shumin Xiao\**

Dr. C. Zhang, Dr. Y. H. Wang, Dr. Y. S. Gao, Dr. Y. B. Fan, Dr. C. Huang, Dr. N. Zhang, Prof. S. M. Xiao, Prof. Q. H. Song
State Key Laboratory on Tunable laser Technology, Ministry of Industry and Information Technology Key Lab of Micro-Nano Optoelectronic Information System, Shenzhen Graduate School, Harbin Institute of Technology, Shenzhen, 518055, P. R. China
Collaborative Innovation Center of Extreme Optics, Shanxi University, Taiyuan, 030006, Shanxi, P. R. China
E-mail: (qinghai.song@hit.edu.cn, shumin.xiao@hit.edu.cn)

Keywords: (Metasurface, Lead halide perovskite, Dynamic control, Hologram, Anomalous reflection)

**Abstract**

Lead halide perovskites (MAPbX$_3$) are known to have high refractive index and controllable bandgap, making them attractive for all-dielectric and tunable metasurfaces. Till now, perovskite metasurfaces have only been used in structural colors. More interesting meta-devices with $2\pi$ phase control are still absent. Here we experimentally demonstrate the MAPbX$_3$ perovskite based metasurfaces with a complete control of phase shift in a reflection mode. By utilizing MAPbBr$_3$ cut-wires as meta-atoms on a ground metal film, we find that the MAPbBr$_3$ perovskite metasurface can produce full phase control from 0 to $2\pi$ and high reflection efficiency simultaneously. Consequently, high-efficiency polarization conversion, anomalous reflection and meta-hologram have been successfully produced. Most interestingly, the bandgap of MAPbX$_3$ perovskite can be post-synthetically and reversibly tuned via anion exchange, providing a new approach to dynamically control of the all-dielectric meta-devices with novel function such as anomalous reflection and hologram et al.

## 1. Introduction

Metasurface is a type of novel two-dimensional system that is composed of subwavelength scale antennas [1–8]. Compared with their uniform counterparts, metasurfaces can accurately control the amplitude, phase, and polarization of the transmitted or reflected light [9-12]. In past few years, a large number of unique functionalities and optical components such as flat lens, holograms, and gratings have been demonstrated in plasmonic metasurfaces and high refractive index dielectric (Si, Ge) metasurfaces [12-15]. Recently, in order to overcome the intrinsic losses, transparent materials such as titanium dioxides (TiO$_2$), silicon nitrides (Si$_3$N$_4$), and gallium nitride (GaN) have recently been applied to replace silicon, germanium, and plasmonic materials for the realization of the meta-devices in the visible spectrum [16-21]. Functional devices including metalens, color printer, hologram, and even achromatic lens have also been successfully demonstrated [16-24]. However, due to their relatively stable properties, the optical performances of such meta-devices are typically fixed and hard to be post-fabrication tuned once they are fabricated as their counterparts made of plasmonic or phase-transition material, e.g. Mg, Pb, VO$_2$, ITO et al [25-28]. Although infiltrating different solutions can partially tune the optical characteristics, the reduced refractive index contrast significantly affect their abilities in fully controlling the phase shift from $0-2\pi$ [29, 30].

Solution-processed lead halide perovskites (MAPbX$_3$, X = I, Cl, Br, and their mixtures) have recently emerged as a class of promising semiconductors for cost-effective optoelectronic devices [31, 32]. Due to the long carrier diffusion length, low defect density, high carrier mobility, and widely tunable bandgap, the photon conversion efficiency (PCE) of lead halide perovskite based solar cell has been dramatically increased from 3.8 % to 23.1 % in a few years [33]. Perovskite based photodetectors, light-emitting diodes, and microlasers have also been rapidly developed simultaneously [34-42]. Recently, lead halide perovskites started to show their potential in metasurfaces and metamaterials [43-45]. Their refractive indices (n ~ 2.2 – 2.55) are comparable to or much larger than conventional transparent materials [46]. And a series of nanofabrication techniques have been developed to convert the solution-processed materials into perovskite nano-structures, e.g. focused ion beam milling (FIB), electron beam lithography (EBL), nano-imprinting, and inductively coupled plasma (ICP) etching [47-49]. Up to now, perovskite based metasurfaces have been successfully fabricated for high-resolution color and dynamic nano-printings [43-45]. However, the more challenging experiments such as complete control of phase shift from 0 to $2\pi$ and functional devices such as metalens and meta-hologram are still absent. Importantly, the intrinsic advantage in in-situ control of material properties has not been fully exploited in perovskite metasurfaces. Herein, we report the MAPbX$_3$ perovskite metasurfaces with complete phase control and explore their potential in dynamic meta-devices for the first time.

## 2. Results and discussions

### 2.1 Synthesis and characterization of lead halide perovskite film

The whole experiment started with a MAPbBr$_3$ film on a gold film, separated by a SiO$_2$ spacer. Basically, a 100 nm gold reflection mirror and a 20 nm SiO$_2$ protecting layer were deposited onto a glass substrate via electron beam (E-beam) evaporation. The deposition rates were 0.04 nm /s and 0.05 nm /s, respectively. The MAPbBr$_3$ film was prepared by spin-coating a solution containing PbBr$_2$ and MABr in a mixture solvent of N,N-dimethylformamide (DMF) and dimethyl sulfoxide (DMSO) (see **Figure 1(a)** and Experimental section) [50]. Here the SiO$_2$ spacer was applied to protect the gold film and to improve the quality of MAPbBr$_3$ film. The top-view scanning electron microscope (SEM) image (inset in **Figure 1(b)**) shows that the synthesized perovskite film is quite uniform and no obvious pinholes can be observed. The grain sizes are on the order of 100 nm. The root mean square value of surface roughness, characterized with atomic force microscope (AFM), is less than 7.2 nm (see supplemental information). **Figure 1(b)** shows the X-ray diffraction (XRD) spectrum of the film. Four obvious peaks can be clearly seen at 15.3$^\circ$, 30.5$^\circ$, 46.2$^\circ$, and 62.9$^\circ$, corresponding to (001), (002), (003), and (004) planes of cubic phase MAPbBr$_3$ well [51]. From the absorption spectrum (**Figure 1(c)**), it is easy to know that the bandgap of MAPbBr$_3$ perovskite is around 2.25 eV. The refractive index (n) and extinction coefficient (k) of MAPbBr$_3$ were also measured and plotted in **Figure 1(d)**. The refractive index of MAPbBr$_3$ is well above 2.1 in the entire visible spectrum (400 nm – 700 nm).

## 2.2 Perovskite metasurfaces based polarization conversion, anomalous reflection, and hologram

The high refractive index, relatively large bandgap, and good quality make MAPbBr$_3$ perovskite films very promising for applications in all-dielectric nano-photonics, especially for the spectral range below the bandgap. Metasurface based broad-band and high-efficiency polarization converter is one good example [52]. Here we take the cut-wire based metasurface as an example to illustrate this possibility. As the SEM image and schematic picture shown in **Figure 2(a)** and **Figure 2(b)**, the length and width of each cut-wire are a = 160 nm and b = 295 nm, respectively. The periods in both directions are 510 nm and the thickness of perovskite cut-wire, measured by profilometry, is 300 nm. Such kinds of perovskite cut-wires are large enough to support resonances along both of long and short axes [53, 54]. For incident light with polarization along x or y directions, it can be decomposed into two perpendicular components along the long and short axes (see **Figure 2(b)**). As the resonances along two axes are different (see supplemental information), the phase retardation between two components are generated. For the current cut-wires, the phase retardation has been optimized to π for a wide spectral range (**Figure 2(c)**). In this sense, the direction of one component shall be reversed and thus the polarization of reflected light is rotated 90° (see right panel in **Figure 2(b)**). The dashed line in **Figure 2(d)** shows the numerically calculated reflection spectrum with cross polarization. The cross-polarized reflection remains above 90 % in a wide range from 540 nm to 680 nm, consistent with the phase retardation in **Figure 2(c)** very well.

Then the MAPbBr$_3$ perovskite metasurface was created by patterning the perovskite film into cut-wires via E-beam lithography and reactive ion etching (see details in Experimental section). The SEM image of the perovskite metasurface (see **Figure 2(a)**) shows that all the structural parameters follow the design very well and no residuals can be observed after the ICP etching. The optical properties of perovskite metasurface have been examined with a white-light source and a CCD camera coupled spectrometer (see Experimental section). The experimental results are summarized as solid line in **Figure 2(d)**. Similar to the numerical calculations, a high-efficiency and broadband polarization reflection spectrum for cross-polarization has been successfully generated with the MAPbBr$_3$ perovskite metasurface. The experimentally recorded absolute efficiency ($\eta=|r_{cross}|^2/I_{incident}$) was around 45 %, which was slightly lower than the numerical calculation. From the high-resolution SEM image (see supplemental information), we know that fabricated cut-wires still have relatively rough boundaries, which are very similar to the initial grain size in **Figure 1(b)**. This might be caused by different etching speed along different crystal directions in MAPbBr$_3$ perovskites and it can be removed if single crystalline film is applied. However, if we define the polarization conversion efficiency following the previous reports ($\eta=|r_{cross}|^2/(|r_{cross}|^2+|r_{co}|^2)$), the experimentally recorded value was above 80 % from 525 nm to 650 nm (see supplemental information), which is comparable to the recent literatures with silicon and TiO$_2$ [52]. Thus we confirm that lead halide perovskite is a good candidate for all-dielectric meta-devices and their performances can be further improved if single crystalline lead halide perovskite is applied. We note that the phase retardation is determined by the different resonances along long and short axes. The sizes and shapes of cut-wires must be precisely controlled. In this sense, while some other techniques can also produce perovskite nanoparticles with Mie resonances, their random sizes and shapes exclude them from the applications in perovskite metasurfaces.

In additional to the phase retardation, the perovskite cut-wires can also produce phase shift between incident light and the reflected light. Interestingly, this kind of phase shift is found to be dependent on the structural parameters. The colormaps in **Figure 3(a)** and **Figure 3(b)** summarize the numerically calculated polarization conversion efficiency (here and below we use the absolute efficiency) and the corresponding phase shift as a function of a and b, respectively. A region with high reflectance ($|r_{cross}|^2$ ~65 %-80 %) can be clearly seen in **Figure 3(a)**. Within this region, it is easy to select four cut-wires (see open circles in **Figure 3(b)** and schematic picture in **Figure 3(c)**) to provide an incremental phase shift of π/4 for the cross-polarized reflected light. An additional π phase shift can be attained by rotating the cut-wire 90º, simply realizing the full 2π coverage and high polarization conversion efficiency simultaneously.

In order to validate the 2π full coverage of phase shift, we have fabricated a metasurface composed of eight cut-wires in **Figure 3(c)** with the same nanofabrication technique as the above. **Figure 4(a)** shows the top-view SEM image of the perovskite metasurface. We can see that eight resonators are placed one another forming a linear phase gradient over a supercell with length L=4.08 µm. The numerical calculation in **Figure 4(b)** shows that the light is anomalously reflected when it illuminates the metasurface at normal incidence. The anomalous reflection angle r=8.91º in **Figure 4(c)** matches the equation derived from generalized Snell's law $\theta_r = \sin^{-1}\left[\left(\lambda_0/nL\right) + \sin\left(\theta_i\right)\right]$, here i =0º is the incident angle. The corresponding experimental results are summarized in **Figure 4(d)**. When the perovskite metasurface was illuminated with a x-polarized He-Ne laser at normal incidence, two reflected laser spots have been experimentally recorded (see the setup in methods and supplemental information). The co-polarized one was the normal reflection, whereas the cross-polarized one corresponded to the anomalous reflection. Here the recorded coefficient of anomalous reflection is about 46 %, consistent with the polarization conversion in **Figure 2** well. Interestingly, the anomalous reflection here also clearly demonstrates the potential of perovskites in meta-lens. Similar to the polarization conversion, the sizes and shapes of cut-wires must be very close to the designs in **Figure 3(c)**. The deviations in cu-wires will degrade or smear out the designed function.

The full coverage of phase shift is also important for the computer-generated hologram (CGH). Based on the selected cut-wires in **Figure 3(c)**, we have designed a cross-polarized hologram with the Gerchberg-Saxton algorithm (see detail phase distribution in supplemental information). The schematic picture is depicted in **Figure 5(a)**. When a He-Ne laser is illuminated onto the sample at normal incidence, an off-axis and cross-polarized holographic image of "HIT" (the abbreviation of our university name, Harbin Institute of Technology) has been generated. The phase distribution of hologram, which is optimized by considering the polarization conversion, signal-to-noise ratio, and uniformity, have been calculated and experimentally endowed to perovskite cut-wires via nanofabrication. The total size of perovskite metasurface is 410 µm× 410 µm. The enlarged top-view SEM image is shown in **Figure 5(b)**, where eight types of cut-wires in **Figure 3(c)** have been fully utilized. Then the metasurface was illuminated with a x-polarized He-Ne laser and its y-polarized reflection image was recorded by a CCD camera. As shown in **Figure 5(c)**, a bright "HIT" can be clearly seen, consistent with the numerical simulation well. The efficiency of anomalous reflection, which is defined as $\eta_{\text{hologram}} = I_{\text{hologram}}/I_{\text{incident}}$, was about 22.02 %. This value is already much higher than the values of plasmonic metasurface or Si metasurface based holograms.

### 2.3 Dynamic anomalous reflection and hologram

From the above results, we know that the lead halide perovskites can be utilized as functional meta-devices and their performances are as good as conventional metasurfaces. However, the most intriguing properties of lead halide perovskites have not been fully exploited yet. Compared with the conventional semiconductors, the optical properties such as bandgap of lead halide perovskite are post-synthetically controllable with many techniques, making dielectric metasurfaces dynamically and reversibly controllable as their plasmonic counterparts as well. Here the perovskites are tuned with the anion exchange in a chemical vapor deposition (CVD) tube [47], (see details in Experimental section and supplemental information). **Figure 6(a)** shows the experimentally recorded XRD spectrum after the anion exchange with MAI vapor. The peaks at 15.3º, 30.5º, 46.2º, and 62.9º in **Figure 1(b)** all disappeared and a series of new peaks emerged at 14.58°, 28.94°, 43.76°, and 59.42°, which could be indexed to the crystal planes of MAPbI$_3$ perovskites. Meanwhile, the bandedge of perovskite film also moved from 2.38 eV to 1.6 eV (see **Figure 6(b)**), clearly confirming the complete transition from MAPbBr$_3$ perovskite to MAPbI$_3$ perovskite. The solid and dashed lines in **Figure 6(c)** show the recorded n and k. After the anion exchange, the refractive index still remained at relatively large value, the light extinction ration k significantly increased. In principle, the increased absorption can simply erase the designed functions of metasurface in the visible spectrum. In this sense, considering the precise and reversibly control of anion exchange, the perovskite based metasurfaces are nice candidates for all-dielectric dynamic meta-devices.

Based on the n and k in **Figure 6(c)**, we first tested the perovskite metasurface based anomalous reflection. The numerically calculated result is plotted as solid line in **Figure 6(d)**, where the coefficient for MAPbBr$_3$ perovskite is plotted as dashed line for a direct comparison. While the angles of anomalous reflection are very close, the anomalous reflectance drastically reduced to < 1/10 of initial value when the MAPbBr$_3$ perovskites were transferred to MAPbI$_3$ perovskites. This kind of transition has also been experimentally verified. The steps in **Figure 6(e)** show the reversible transition between "ON" and "OFF" status of the MAPbX$_3$ perovskite metasurface. After the transition to MAPbI$_3$ perovskite, the absorption at 633 nm was greatly enhanced, whereas the refractive index was close [38]. As a result, the phase shifts were well preserved but the intensity was significantly reduced from 46 % to 7 %. This switching off process can also be seen from the recorded photographs. As the insets shown in **Figure 6(e)**, The position of anomalous reflection spot was still kept at r=8.91°, and its intensity was barely seen. Interestingly, the transition between MAPbBr$_3$ perovskite and MAPbI$_3$ perovskite is reversible. In our experiment, the efficiency of anomalous reflection has been fully recovered after converting the MAPbI$_3$ back to MAPbBr$_3$ perovskites. Therefore, the experimental results in **Figure 6(e)** clearly confirmed the potential of lead halide perovskite in dynamic meta-devices, making the dielectric metasurface comparable to their plasmonic counterparts for the first time.

The dynamic control technique via anion exchange is also applicable to the perovskite based CGH. Similar to the anomalous reflection, the dominate changes happened in the light extinction coefficient (k) after the anion exchange. In this sense, the designed image shall be preserved and the intensity is reduced from > 20% to <2% when the MAPbBr$_3$ was transferred to MAPbI$_3$. This is exactly what we have observed in experiment. As shown in **Figure 6(f)**, the holographic image "HIT" has been almost fully erased and only the diffracted light from central point can be seen. Owing to the unique property of MAPbX$_3$ perovskites, the perovskite based meta-hologram has also been switched from "OFF" state to "ON" state by converting the MAPbI$_3$ perovskite back to MAPbBr$_3$ perovskite, making the perovskite meta-hologram dynamically and reversibly switchable. Note that the "OFF" state absorbed most of incident laser. However, the perovskite metasurface only produced photoluminescence instead of lasing under continuous excitation at room temperature [47]. As a result, the near-infrared photoluminescence didn't change the hologram patterns in **Figure 6 (f)**.

Till now, based on the experimental results on both of dynamic anomalous reflection and dynamic hologram, we confirm that lead halide perovskite can be a nice platform to fill the gap between dielectric metasurface with complete phase control and dynamic functions, which are only attainable in plasmonic metasurface very recently [27]. The "ON-state" is operating in the transparent spectral range and experience very low absorption loss. Meanwhile, according to the recent developments, the transition time can be as short as tens of seconds or below one second if the other anion exchange techniques are applied [55-57]. Remarkably, lead halide perovskites have shown exceptional nonlinear properties, making the nonlinear meta-devices attainable (see supplemental information). All these performances simply make perovskite metasurfaces superior than the previous reports.

## 3. Conclusion

In summary, we have experimentally demonstrated the MAPbX$_3$ perovskite based metasurfaces with a complete control of phase shift in a reflection mode. By patterning MAPbBr$_3$ cut-wires as meta-atoms on a ground metal film, we find that the MAPbBr$_3$ perovskite metasurface can produce full phase control from 0 to $2\pi$ and high reflection efficiency simultaneously. Consequently, high-efficiency polarization conversion (~45 % for absolute efficiency, ~ 80 % for conventional definition), anomalous reflection (~45 %) and hologram (~22 %) have been experimentally realized. These values can be further improved by using single-crystalline lead halide perovskite microplates or film to replace the polycrystalline film. Importantly, by post-synthetically and reversibly control the optical properties of MAPbX$_3$, we have demonstrated that the metasurface based anomalous reflection and hologram can be simply switched "ON" and "OFF" via anion exchange. Our experiment is the first experimental realization of dynamic all-dielectric metasurface in the visible spectrum. It can pave a new route to the dynamic metasurfaces and dynamic meta-devices.

## 4. Experimental Section

*Deposition of Au and SiO$_2$ films*: The Au and SiO$_2$ films were deposited onto glass substrate with E-beam evaporation (SKE_A_75). The base vacuum pressure was $2 \times 10^{-7}$ Torr. The deposition rates of Au and SiO$_2$ were 0.4 Å/s, and 0.5 Å/s, respectively. After the deposition of SiO$_2$, the substrate was treated with oxygen plasma (MYCRO, FEMTO6SA) for 12 seconds to achieve the hydrophilic surface.

*Preparation of Perovskite Films:* The MAPbX$_3$ perovskite film was prepared by spin-coating the MAPbX$_3$ precursor onto a substrate coated with Au and SiO$_2$. The MAPbX$_3$ precursor was obtained by dissolving PbBr$_2$ and CH$_3$NH$_3$Br (99.999 %, Shanghai MaterWin New Materials co.) with a 1:1 molar ratio in dimethylsulfoxide. The solution was stirred for 6 h and filtrated by 0.2 um polytetrafluoroethylene (PTFE) syringes before use. To get the perovskite film with a thickness of 220 nm, the 35 ul 1.2M $CH_3NH_3PbBr_3$ precursor was spin-coated onto the Au layer at 4000 r.p.m. for 90s. At the 23th second of spinning, 70 µL of chlorobenzene was quickly dropped on the film to promote formation of uniform and dense lead halide perovskite film. And for the thickness of 300nm, that is 1.4 M precursor at 3500 r.p.m. for 90s and chlorobenzene was dropped at 27th second. Notably, the above prepared process was conducted in the glovebox with $Ar_2$ gas at room temperature.

*Fabrication of Perovskite Metasurfaces:* The $MAPbBr_3$ metasurfaces includes an electron-beam lithography and an inductively coupled plasma etching. 400 nm electron-beam resist (ZEP-520A) was spin-coated onto the perovskite film and then patterned by electron beam writer (Raith E-line) with a dose 90 C/cm2 under an acceleration voltage 30 kV. After developing in N50, the pattern of metasurface was generated in the E-beam resist. The pattern was transferred to $MAPbBr_3$ perovskite with an etching process (Oxford Instruments, PlasmaPro ICP180). The vacuum degree was $10^{-9}$. The ICP power was 600 W, and the RF power was 150 W. Under this condition, the $MAPbBr_3$ film was etched by chlorine gas with 5 sccm flow rate. $C_4F_8$ with 10 sccm flow rate was used as protective gas.

*Optical Characterization:* The anomalous reflection and hologram of perovskite metasurface were characterized with a homemade optical setup. The details are shown as **Figure S3** in supplemental information. During the optical measurements, both of co-polarized light and cross-polarized light were characterized and their intensities were $I_N$ and $I_A$, respectively. Next, the intensity from a pure Au mirror was measured as the total input power ($I_{Au}$). Thus, the anomalous reflectance η ($\eta = I_A / I_{Au}$) was calculated. Similar method is also used for the characterization of hologram image.

*Conversion between MAPbBr3 and MAPbI3:* The $MAPbBr_3$ perovskite was converted to $MAPbI_3$ perovskite in a CVD tube. During the whole vapor conversion process, the MAI powder was placed at the center of the CVD furnace while the $MAPbBr_3$ patterns which have already prepared on the substrate were mounted downstream of the apparatus. The central heating zone was increased to 125 °C (8 °C/min heating rate, 10 mins as buffer) under low-pressure conditions (40−50 Torr) and maintained from 25 min to 2 h. Ar and H2 were used as carrier gases with flow rates of 35 and 15 sccm, respectively. The furnace was then naturally cooled to room temperature and MAPbI$_3$ microplates have been obtained. The reversed conversion process is similar.

*Numerical Simulations:* Full-wave numerical simulations of near fields were carried out by using the commercial finite-element method solver COMSOL Multiphysics. The calculated geometries were taken from experimentally obtained SEM images. The dielectric permittivity of the silica substrate was 2.25, whereas the optical properties of the perovskite film have a strong dispersion and has been considered (See **Figure S1**). The maximum value of permittivity of the perovskite was about 6.5, which is much larger as compared with values for the substrate and air, providing a high optical contrast for Mie resonance excitation.

## Acknowledgements

This research is finally supported by National Natural Science Foundation of China under the grant No. 91850204 and Shenzhen Fundamental research projects under the grant No. JCYJ20160427183259083.

(Supporting Information is available online from Wiley InterScience or from the author).

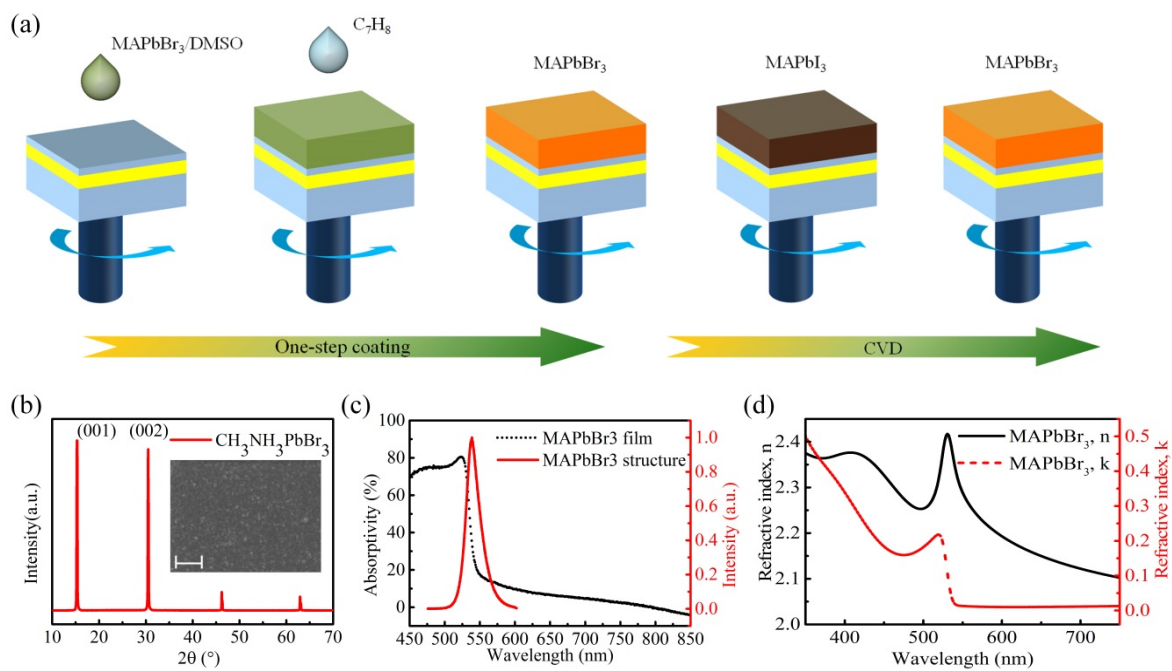

**Figure 1. The lead halide perovskite (MAPbX₃) films.** (**a**) The schematic picture of synthesis process. (**b**) The XRD spectra for the as-grown MAPbBr₃ perovskite film. The inset is the top-view SEM imageof the MAPbBr₃ film.. The scale bar is 1 μm. (**c**) and (**d**) show the corresponding absorption and fluorescence spectra and *n, k* of MAPbBr₃ perovskite films.

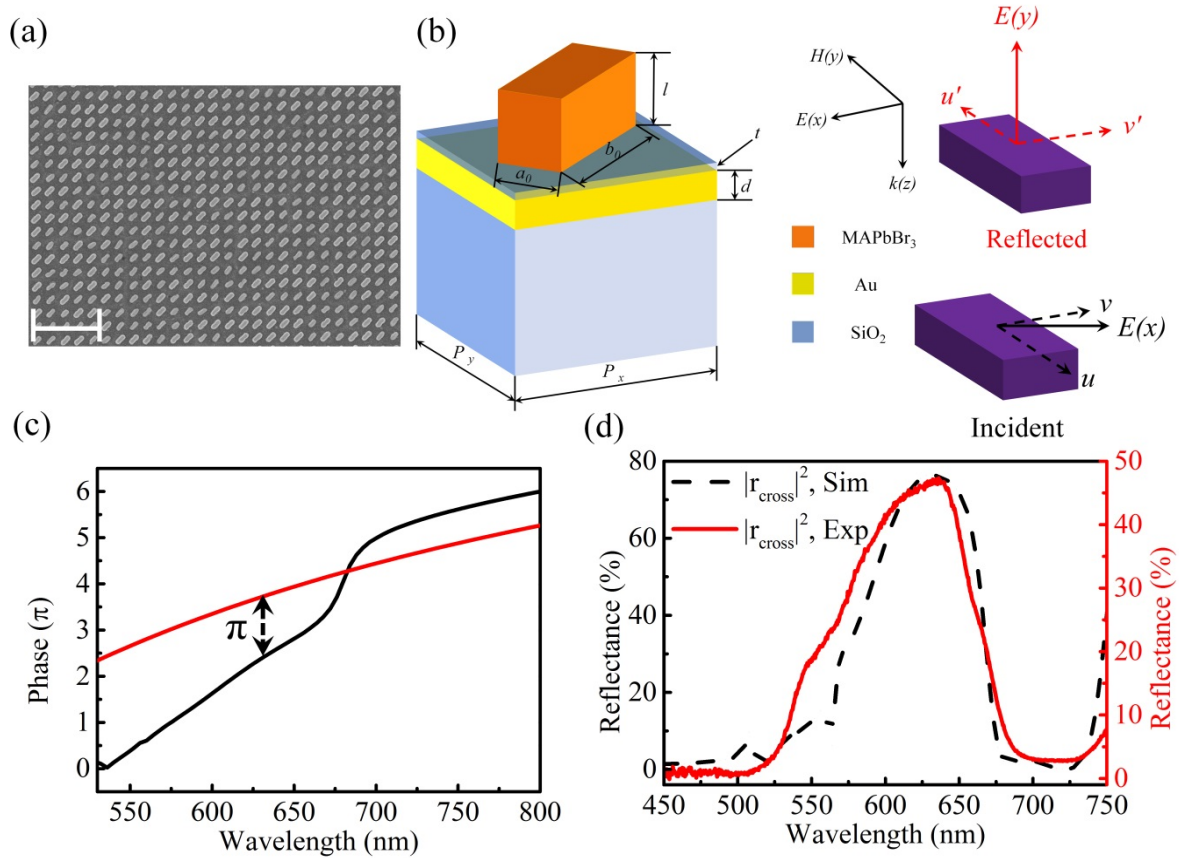

**Figure 2. The MAPbBr₃ metasurface based polarization converter.** (**a**) The tilt-view SEM image of the perovskite metasurface. The scale bar is 2 μm. (**b**) The illustration of polarization rotation via phase retardation. (**c**) The phase retardation between two components along long and short axes. (**d**) The numerically calculated and experimentally recorded reflection spectra for cross-polarization.

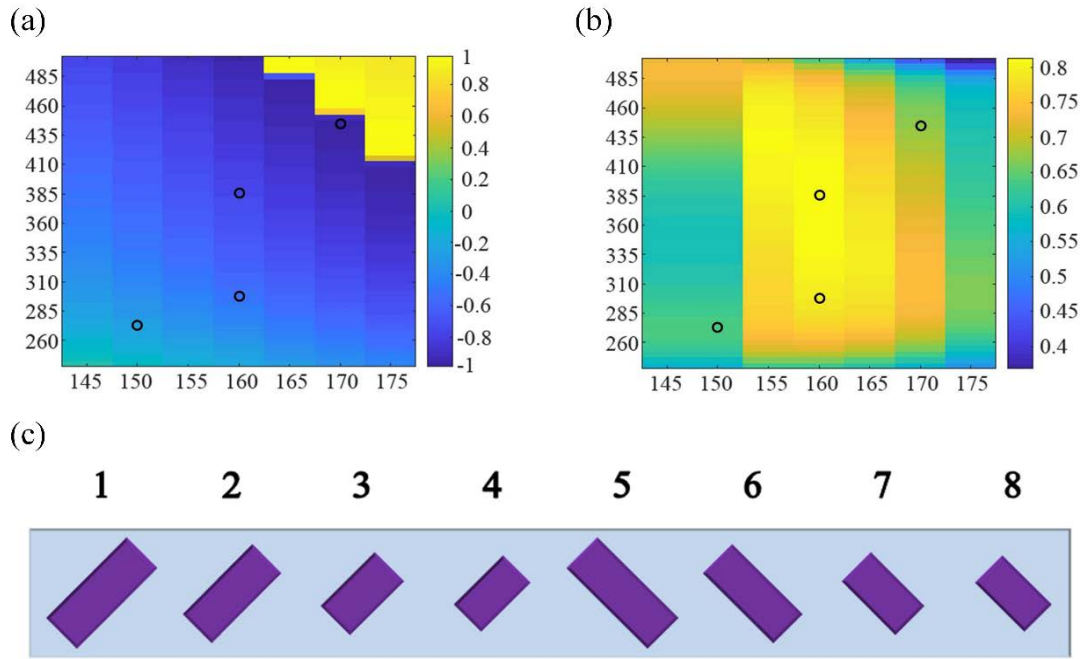

**Figure 3. Numerically simulated cross-polarized reflection.** (**a**) and (**b**) are the reflectance and phase of cut-wires with varying size parameter *a* and *b*. Here the incident laser is fixed at 632.8 nm for He-Ne laser. (**c**) The selected cut-wires in (**b**) for realization of 0-2π phase control.

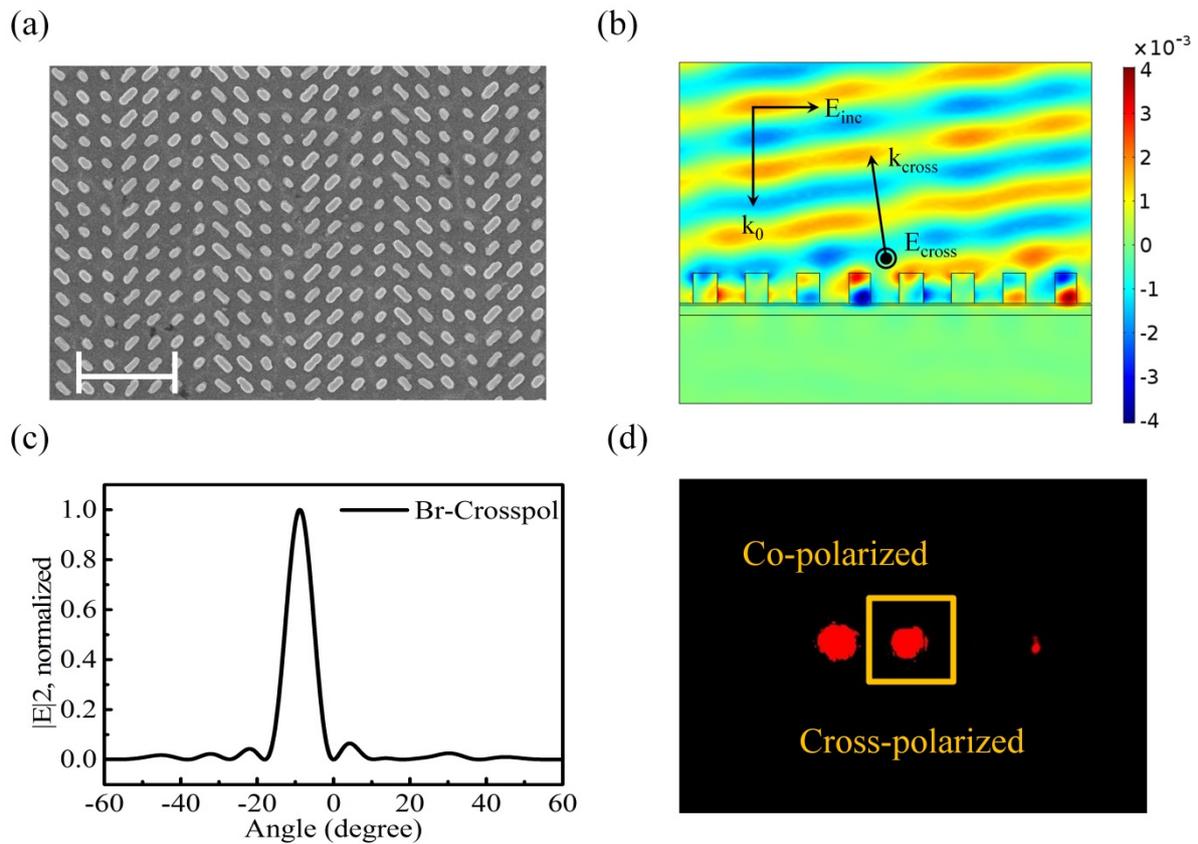

**Figure 4. Dynamic anomalous reflection.** (**a**) The top-view SEM image of MAPbBr₃ metasurface. The scale bar is 2 μm. (**b**) The field distributions of MAPbX₃ perovskite metasurface illuminated with a He-Ne laser at normal incidence. (**c**) The calculated far field angular distributions of anomalous reflection for MAPbBr₃ metasurface. (**d**) The corresponding experiment results of anomalous reflection.

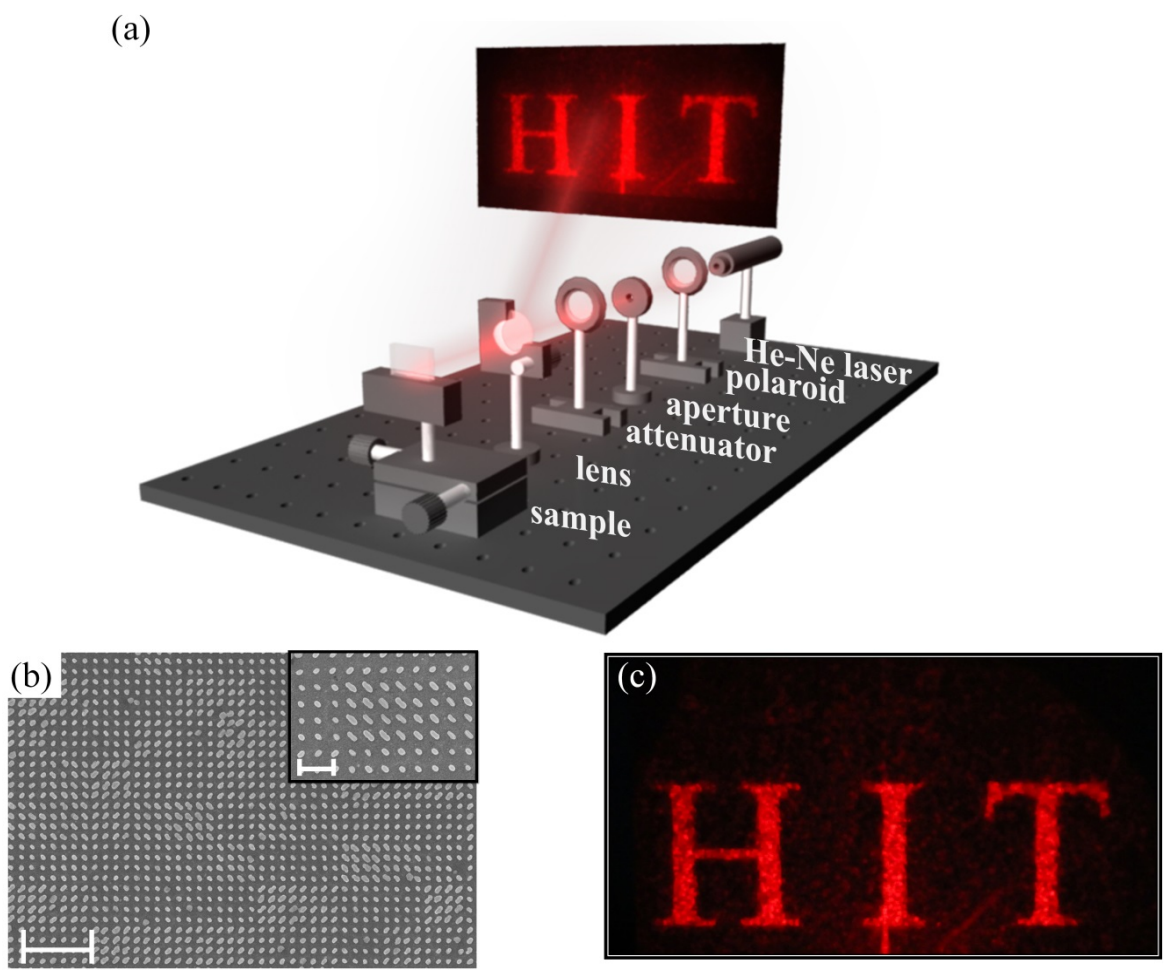

**Figure 5. The perovskite holographic image.** (**a**) The schematic picture of the perovskite metasurface based hologram in a reflection mode. (**b**) The top-view SEM image of the MAPbBr₃ perovskite based metasurface. The scale bar is 3 μm. The inset shows the high resolution SEM of perovskite metasurface with 1 μm scale bar. (**c**) is the experimentally recorded holographic image "HIT" from MAPbBr₃ perovskite metasurface.

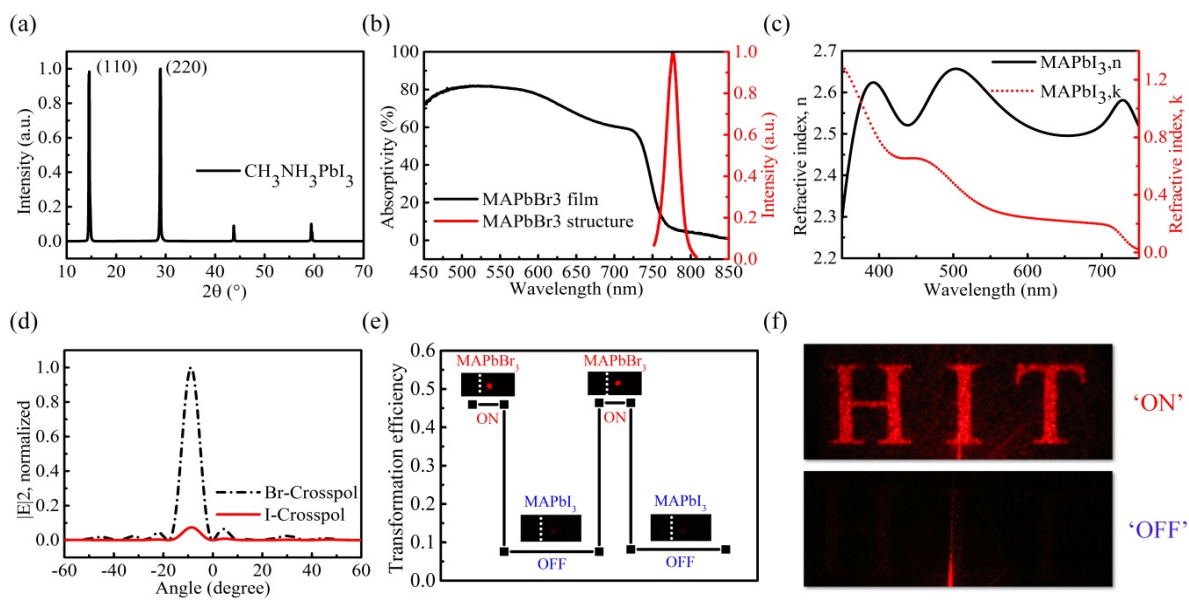

**Figure 6. Dynamic holographic image.** (a) The XRD spectra for the as-grown MAPbI₃ perovskite film. (**b**) and (**c**) show the corresponding absorption and fluorescence spectra and *n, k* of MAPbI₃ perovskite films. (**d**) The calculated far field angular distributions of anomalous reflection from MAPbBr₃ metasurface (dashed line) and MAPbI₃ metasurface (solid line). (**e**) The experimentally recorded "ON" and "OFF" of the anomalous reflection. The insets are their corresponding reflection images. (f) The experimentally recorded holographic image "HIT" in with "ON" and "OFF" from MAPbBr₃ perovskite metasurface and coverted MAPbI₃ perovskte metasurface, respectivity.

**Supporting Information**

**1. The experimental setups.**

**1.1 The CVD conversion setup and process**

In case of lead halide perovskite based dynamic meta-devices, the transition between MAPbBr$_3$ perovskite and MAPbI$_3$ perovskite played an essential role. In additional to the description in the methods, here we utilize the schematic pictures to illustrate the corresponding processes.

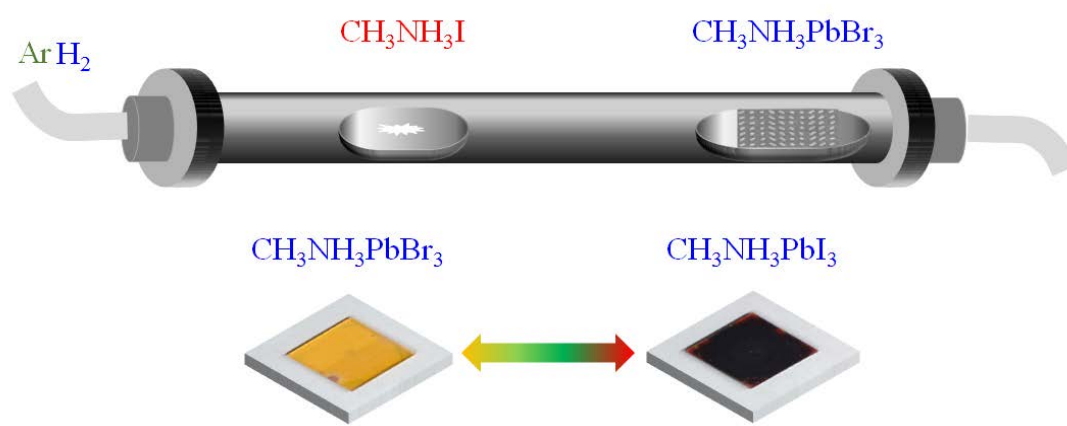

**Figure S1**. The conversion from MAPbBr$_3$ perovskite to MAPbI$_3$ perovskite

As shown in **Figure S1**, the MAPbBr$_3$ perovskite was converted to MAPbI$_3$ perovskite in a CVD tube. During the whole vapor conversion process, the MAI powder was placed at the center of the CVD furnace. The MAPbBr$_3$ metasurfaces were mounted downstream of the apparatus. The central heating zone was increased to 125 °C (8 °C/min heating rate, 10 mins as buffer) under low-pressure conditions (40−50 Torr) and maintained from 25 min to 2 h. Ar and H$_2$ were used as carrier gases with flow rates of 35 and 15 sccm, respectively. After the conversion, the CVD furnace was naturally cooled down to room temperature.

The transition process from MAPbI$_3$ perovskite back to MAPbBr$_3$ perovskite was also realized with the same CVD furnace. As shown in **Figure S2**, the MAPbI$_3$ patterns on the substrate were mounted downstream of the apparatus and the MABr powder was placed at the center of the CVD furnace. The central heating zone was increased to 120 °C (8 °C/min heating rate, 13 mins as buffer).

## 1.2 The schematic process of nanofabrication

**Figure S2** below shows the schematic picture of the nanofabrication process. Basically, 400 nm electron-beam (E-beam) resist (ZEP520A) was spin-coated onto the perovskite film. The E-beam resist was baked at 150°C for 30 minutes before patterning it with E-beam writer (Raith, E-line). During the pattering, the dose and acceleration voltage were kept at 90 μC/cm² and 30 kV, respectively. Then the patterned film was developed in N50 developer for 1 minute at room temperature. After the developments, the designed patterns were successfully generated in the E-beam resist.

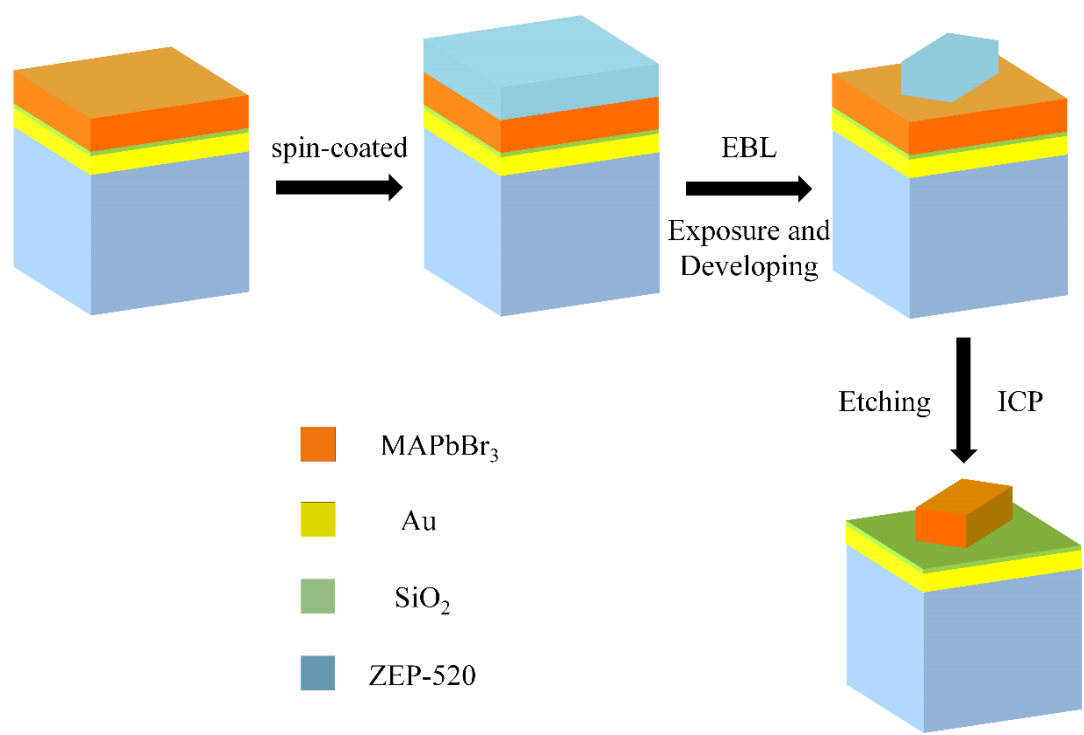

**Figure S2**. The schematic picture of the nano-fabrication process

After the E-beam patterning, it is more essential to transfer the pattern from E-beam resist to perovskite film. In our experiment, this process was realized with reactive ion etching in a inductively coupled plasma (ICP, Oxford Instruments, PlasmaPro ICP 180) etcher. The basic vacuum was reduced to $10^{-9}$. The ICP and RF powers were optimized to 600 W and 150 W, respectively. The cholerine ions (Cl⁻) were employed to etch the perovskite and $C_4F_8$ was applied to protect the sidewall. In our experiment, the flow rates of two gas were 5 sccm and 10 sccm, respectively. The tested etching rate of perovskite was 19.34 nm/s. And the tilt angle or sidewall was ~ 88°, which was good enough for most of photonic applications.

### 1.3 The measurements of reflection spectra of meta-devices, anomalous reflection, and hologram

The setup for optical characterization is plotted in **Figure S3**. Basically, a white light (SLS201L/M) is collimated by an optical lens group and a pin-hole filter. Then the collimated beam passes a polarizer and focused onto the top surface of samples (perovskite films and metasurfaces) via an objective lens (20X, NA = 0.29, focus length l = 20 mm). The intensity and spot size are controlled by the and neutral density filter and diaphragm, respectively. The reflected light is collected by the same objective lens and coupled to a spectrometer (Maya2000) after passing another polarizer. The transmitted light is collected by another optical lens and coupled to the same spectrometer after passing another polarizer.

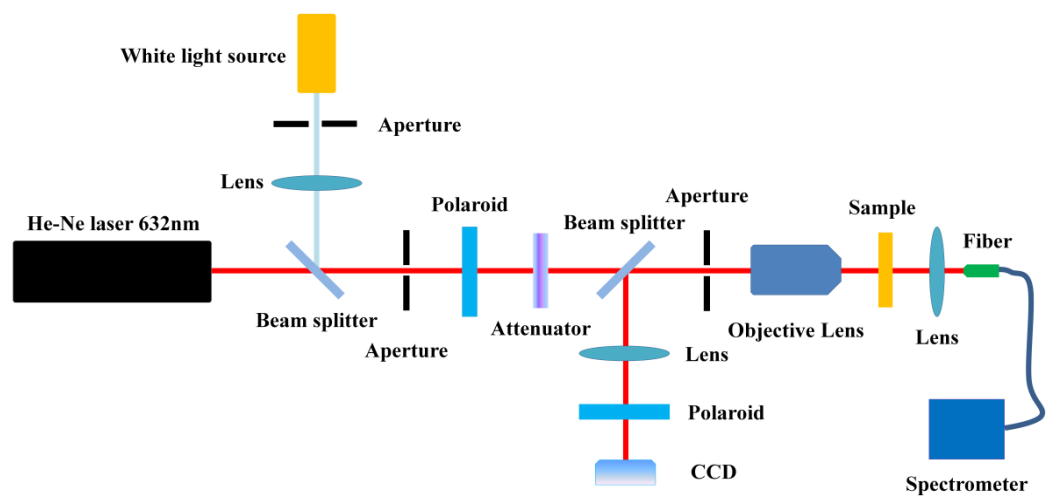

**Figure S3**. The optical setup for measurements of reflection spectra

In case of anomalous reflection and hologram, a He-Ne laser has been added into the setup. Similar to the white light measurement, the He-Ne laser passes a polarizer and incidents on the sample via the objective lens. The reflected light is characterized by a CCD camera after another polarization to character the normal reflection (co-polarization) and the anomalous reflection (cross-polarization). For the hologram measurement, only the cross-polarization has been measured. The image is recorded by a CCD camera. The hologram image is collected by a lens and measured by a power meter.

## 2. The characterization of perovskite films.

For the perovskite metasurfaces, the roughness of perovskite film is quite essential. In this experiment, we have measured the surface roughness with atomic force microscope (AFM, CSPM5500). **Figure S4** shows the experimental results. By mapping a 10 µm×10 µm area, the rout mean square value is as mall as 7.2 nm. This is good enough for conventional photonic devices. In addition, the AFM image shows the polycrystalline nature of perovskite films. The different grains can also be clearly seen in **Figure S4**.

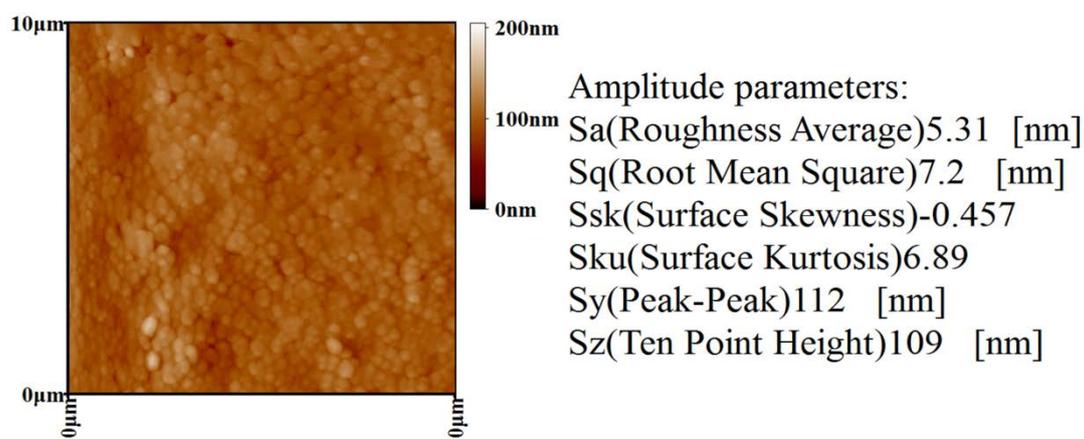

Amplitude parameters:
Sa(Roughness Average)5.31 [nm]
Sq(Root Mean Square)7.2 [nm]
Ssk(Surface Skewness)-0.457
Sku(Surface Kurtosis)6.89
Sy(Peak-Peak)112 [nm]
Sz(Ten Point Height)109 [nm]

**Figure S4**. The AFM image of the perovskite film and corresponding surface roughness analysis

## 3. The characterization of perovskite meta-devices.

### 3.1 The optical characterization of perovskite polarization conversion

For the anomalous reflection, we have numerically computed the reflection spectra of both polarizations along long and short axes of the cut-wires. The results are shown in Figure S5. We can see that both polarizations can be highly reflected in a wide spectral range from 580 nm to 670 nm. Meanwhile, due to two resonances can generate a $\pi$ phase retardation (see main text), the linear polarization along 45 degree can be rotated 90 degree and thus forms the polarization conversion.

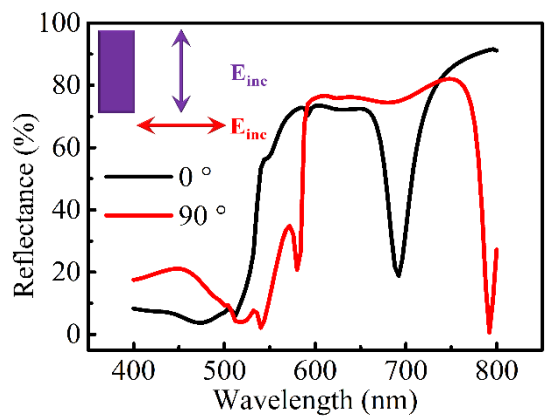

**Figure S5**. The numerically calculated reflection spectra of two polarization along the long (black) and short (red) axes of perovskite cut-wires

The detailed numerical results for a polarization along 45 degree are summarized in Figure S6 below. The red and black curves in **Figure S6 (a)** are the numerically calculated reflection spectra with co-polarization and cross-polarization. It is easy to see that two types of reflection spectra are complement each other. This is caused by the phase retardation. In the spectral range from 580 nm to 670 nm, the phase retardation is around π and thus cross-polarization is formed. By using the same setup, both of the co-polarized and cross-polarized reflection spectra have been experimentally measured. The corresponding results are summarized in **FgiureS6(c).** The whole trends are similar to the numerical calculation except that the spectral ranges blue-shifted. This might be caused by the size deviations of cut-wires during the nanofabrication process. The comparison between experimental results with cross-polarization and co-polarization are also plotted in the main text and **Figure 6(d)**, respectively.

The black line in **Figure S6 (b)** shows the polarization conversion coefficient, which is calculated following the conventional definition ($\eta = |r_{cross}|^2/(|r_{cross}|^2+|r_{co}|^2)$). We can see that the conversion efficiency is close to 100% (note this is relative ratio not the absolute value). The red line is the corresponding experimental results. As mentioned in the main text, the relative coefficient is as high as 80%. Note that the experimental result is on the shorter wavelength range due to the fabrication deviations. Meanwhile, the presence of surface roughness not only reduces the absolute coefficient. The relative coefficient also becomes lower and broader.

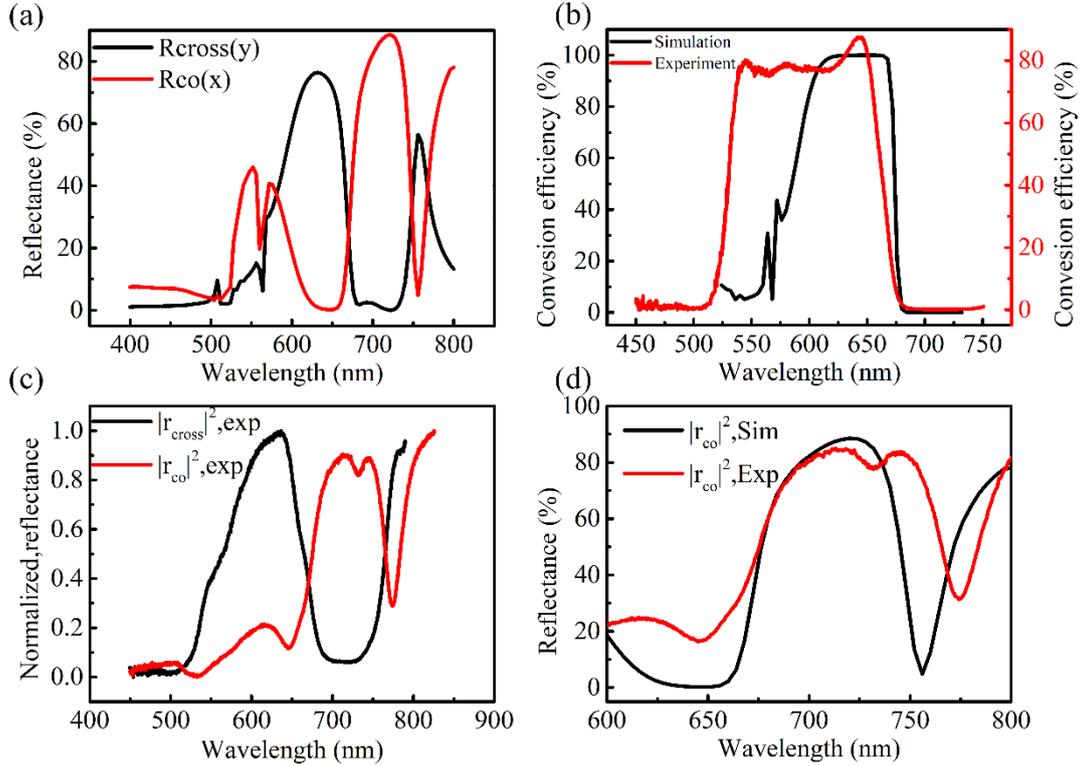

**Figure S6.** (**a**) The reflection spectra of co-polarization (red) and cross-polarization (black). (**b**) The calculated (red) and experimentally recorded (black) polarization conversion efficiencies. (**c**) The experimentally measured reflection spectra of co-polarization (red) and cross-polarization (black). (**d**) The comparison between experimentally measured (red) and numerically calculated (black) reflection spectra.

### 3.2 The high-resolution SEM image of top-down fabricated perovskite nanostructures

In the main text, we have measured the polarization conversion efficiency as high as 45%. Compared with silicon and plasmonic metasurfaces, this value was quite good but was still lower than our numerical calculations. Interestingly, if we look at the ratio between cross-polarized reflection over total reflection, the conversion efficiency can be as high as 80%. This indicates that most of energy loss comes from the reflection at the perovskite metasurface. As the material loss is almost zero in this wavelength region, the losses should be generated by the scattering from surface roughness of cut-wires. This is exactly what we have observed with high resolution SEM images.

(a)                                                    (b)

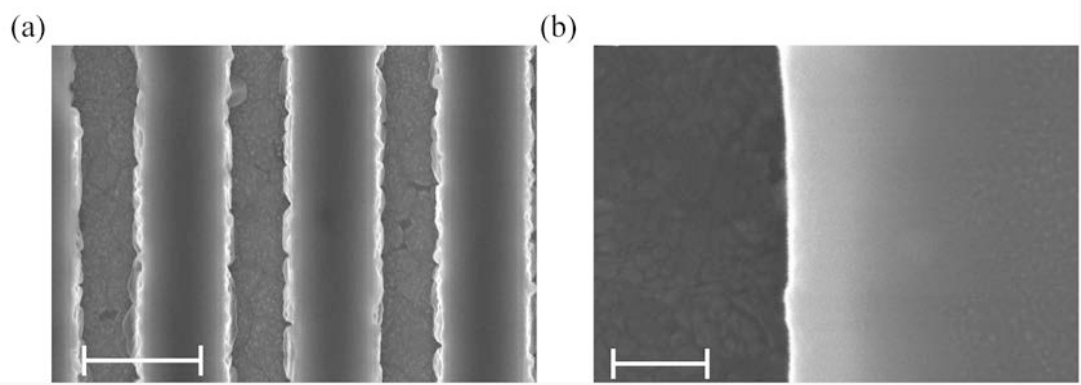

**Figure S7**. (a) The high-resolution top-view SEM image of the top-down fabricated perovskite waveguides by film, Surface roughness can be clearly seen on the sidewalls. The scale bar is 1 µm. (b) The high-resolution top-view SEM image of the top-down fabricated perovskite single crystal, and have a slide edge. The scale bar is 500 nm.

As the cut-wires are too short to fully express the surface roughness, below we take a perovskite waveguide to illustrate this effect. **Figure S7** shows the high-resolution top-view SEM image of the top-down fabricated MAPbBr$_3$ perovskite waveguide. The fabrication process is the same the perovskite metasurface. The bottom of etched region is very clean and the structures of bottom ITO can be clearly seen. This means that the perovskites were fully etched. In case of the side walls, it is clearly to see the roughness along the waveguide. This kind of roughness is quite generic in all metasurfaces and thus can strongly affect the absolute efficiency of polarization conversion. As the roughness is close to the grain sizes in **Figure 1(b)** of main text, we believe this should be generated by the different etching speed along different crystal directions in lead halide perovskite polycrystalline film. In this sense, the perovskite-based devices can be further improved by replacing the polycrystalline film with single-crystalline microplates or films.

**3.3 The design of perovskite anomalous reflection**

In the numerical calculations, we have carefully designed the metasurface for anomalous reflection. As shown in **Figure S8 (a)**, the light incidents on the metasurface at normal direction. Its reflection happens at an angle θ =8.9 degree. In principle, this anomalous reflection can be understood with the general Fresnel law. The corresponding far field angular distribution is plotted in **Figure S8 (b)**.

Interestingly, the field pattern and far field angular distribution of the same metasurface after transition has also been calculated by using the refractive index and light extinction coefficient of MAPbI$_3$ perovskite. The corresponding results are plotted in **Figure S8(c)** and **Figure S8 (d)**. With the same scalebar as **Figure 8(a)** and **(b),** it is easy to see that the intensity of anomalous reflection is significantly reduced even though the anomalous reflection angle is still kept at 8.9 degree. All of these results are consistent with our experimental observations.

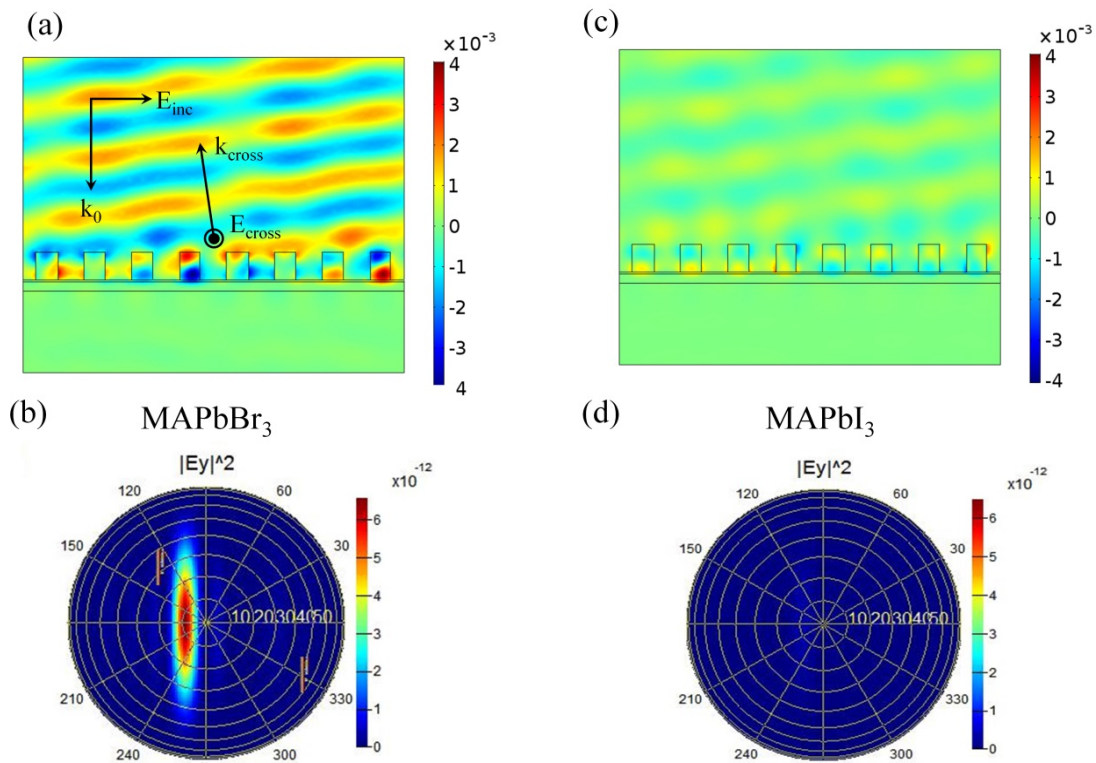

**Figure S8**. The numerically calculated near-field (**a**) and far field (**b**) distributions of anomalous reflection of a MAPbBr$_3$ perovskite metasurface. The corresponding results of MAPbI$_3$ perovskite metasurface.

### 3.4 The design and optical characterization of meta-hologram

The meta-hologram was designed following the Gerchberg-Saxton algorithm. By using the phase information in the main text, a hologram for normal incident light with polarization along x-axis has been designed. The phase distributions obtained for the hologram are shown in **Figure S9 (b)**. The conversion efficiency, signal-to-noise ratio, and uniformity were taken as merit functions for optimization. **Figure S9 (a)** shows the target image and the simulated holographic image, which includes the abbreviation of our university Harbin Institute of Technology (HIT). This is consistent with our experimental results in the main text well.

(a)

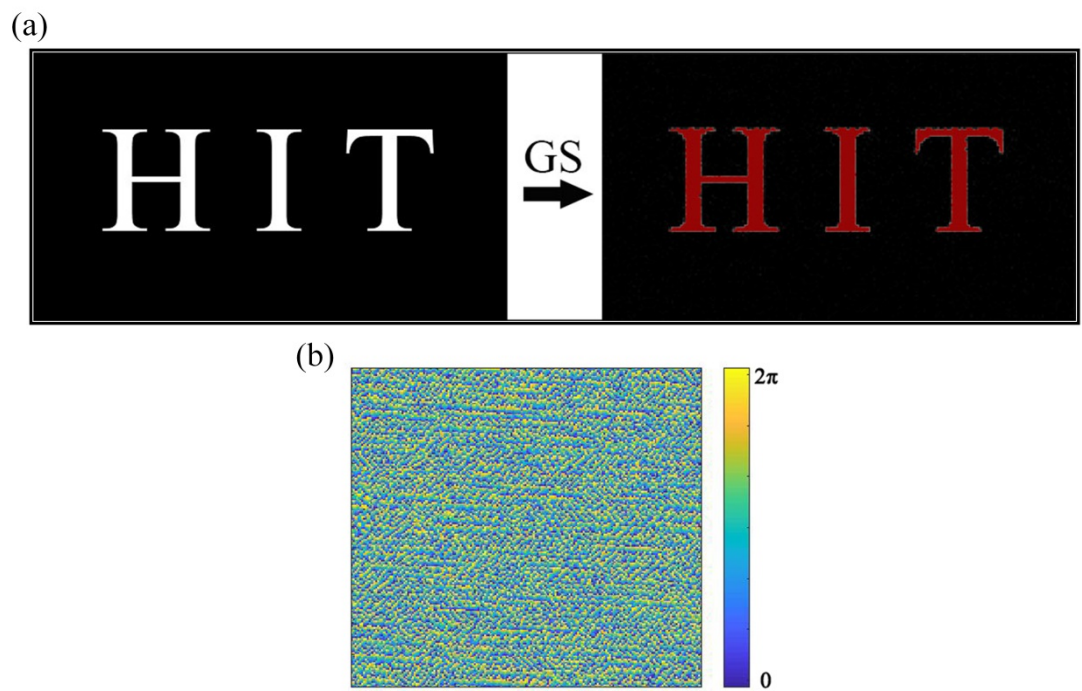

(b)

**Figure S9.** The computed -generated HIT image. The bottom figure is the numerically calculated phase distributions.